\documentclass[aps, reprint,superscriptaddress]{revtex4-1}

\usepackage{amsmath}
\usepackage{amssymb}
\usepackage{mathtools}
\usepackage{graphicx}% Include figure files
\usepackage{dcolumn}% Align table columns on decimal point
\usepackage{bm}% bold math
\usepackage{hyperref}% add hypertext capabilities
\usepackage[mathlines]{lineno}% Enable numbering of text and display math
\usepackage{tabls}

\usepackage{algorithm}
\usepackage[noend]{algpseudocode}

\usepackage{lipsum}

\usepackage[bottom]{footmisc}

\begin{document}

\preprint{APS/123-QED}

%Ratcheting Dynamics on A Funneled Energy Landscape Powers Contraction In Actomyosin-Cross-linked Quasi-Sarcomeric Elements
%\title{The Emergence of Contractility in Non-Muscle Actomyosin: Thermodynamic and Kinetic Principles Governing Dynamics of Elementary Force Dipoles}%\thanks{A footnote to the article title}%

\title{Stochastic Ratcheting on a Funneled Energy Landscape is Necessary for Highly Efficient Contractility of Actomyosin Force Dipoles}

\author{James E. Komianos}
\affiliation{Biophysics Graduate Program, University of Maryland, College Park, MD 20742, USA}
 \affiliation{Department of Chemistry and Biochemistry, University of Maryland, College Park, MD 20742, USA}
 \affiliation{Institute for Physical Science and Technology, University of Maryland, College Park, MD 20742, USA}
\author{Garegin A. Papoian}
 \email{gpapoian@umd.edu}
 \affiliation{Department of Chemistry and Biochemistry, University of Maryland, College Park, MD 20742, USA}
 \affiliation{Institute for Physical Science and Technology, University of Maryland, College Park, MD 20742, USA}

\date{\today}% It is always \today, today,

\begin{abstract}

%In cellular cytoskeletal dynamics, in order to create contractile force within the cell or alter overall network organization, intra- or extra-cellular cues can trigger actin protein fibers to be sheared by myosin II molecular motors - this produces self-organized, cellular-scale morphological changes which can drive biological processes. With cross-linking proteins connecting filamentous structures being essential to the behavior of these networks, recent works have quantified their effect by defining a system connectivity which controls the propagation of contraction between contractile network nodes. 

Current understanding of how contractility emerges in disordered actomyosin networks of non-muscle cells is still largely based on the intuition derived from earlier works on muscle contractility. In addition, in disordered networks, passive cross-linkers have been hypothesized to percolate force chains in the network, hence, establishing large-scale connectivity between local contractile clusters. This view, however, largely overlooks the free energy gain following cross-linker binding, which, even in the absence of active fluctuations, provides a thermodynamic drive towards highly overlapping filamentous states. In this work, we shed light on this phenomenon, showing that passive cross-linkers, when considered in the context of two anti-parallel filaments, generate noticeable contractile forces. However, as binding free energy of cross-linkers is increased, a sharp onset of kinetic arrest follows, greatly diminishing effectiveness of this contractility mechanism, allowing the network to contract only with weakly resisting tensions at its boundary. We have carried out stochastic simulations elucidating this mechanism, followed by a mean-field treatment that predicts how contractile forces asymptotically scale at small and large binding energies, respectively. Furthermore, when considering an active contractile filament pair, based on non-muscle myosin II, we found that the non-processive nature of these motors leads to highly inefficient force generation, due to recoil slippage of the overlap during periods when the motor is dissociated. However, we discovered that passive cross-linkers can serve as a structural ratchet during these unbound motor time spans, resulting in vast force amplification. Our results shed light on the non-equilibrium effects of transiently binding proteins in biological active matter, as observed in the non-muscle actin cytoskeleton, showing that highly efficient contractile force dipoles result from synergy of passive cross-linker and active motor dynamics, via a ratcheting mechanism on a funneled energy landscape.

%

%\begin{description}
%\end{description}
\end{abstract}

\pacs{Valid PACS appear here}% PACS, the Physics and Astronomy
\maketitle

%\tableofcontents

\DeclareRobustCommand{\rchi}{{\mathpalette\irchi\relax}}
\newcommand{\irchi}[2]{\raisebox{\depth}{$#1\chi$}} % inner command, used by \rchi

\section{\label{sec:level1}Introduction}

\begin{figure}[!ht]
\centering
\includegraphics[width=3.4in]{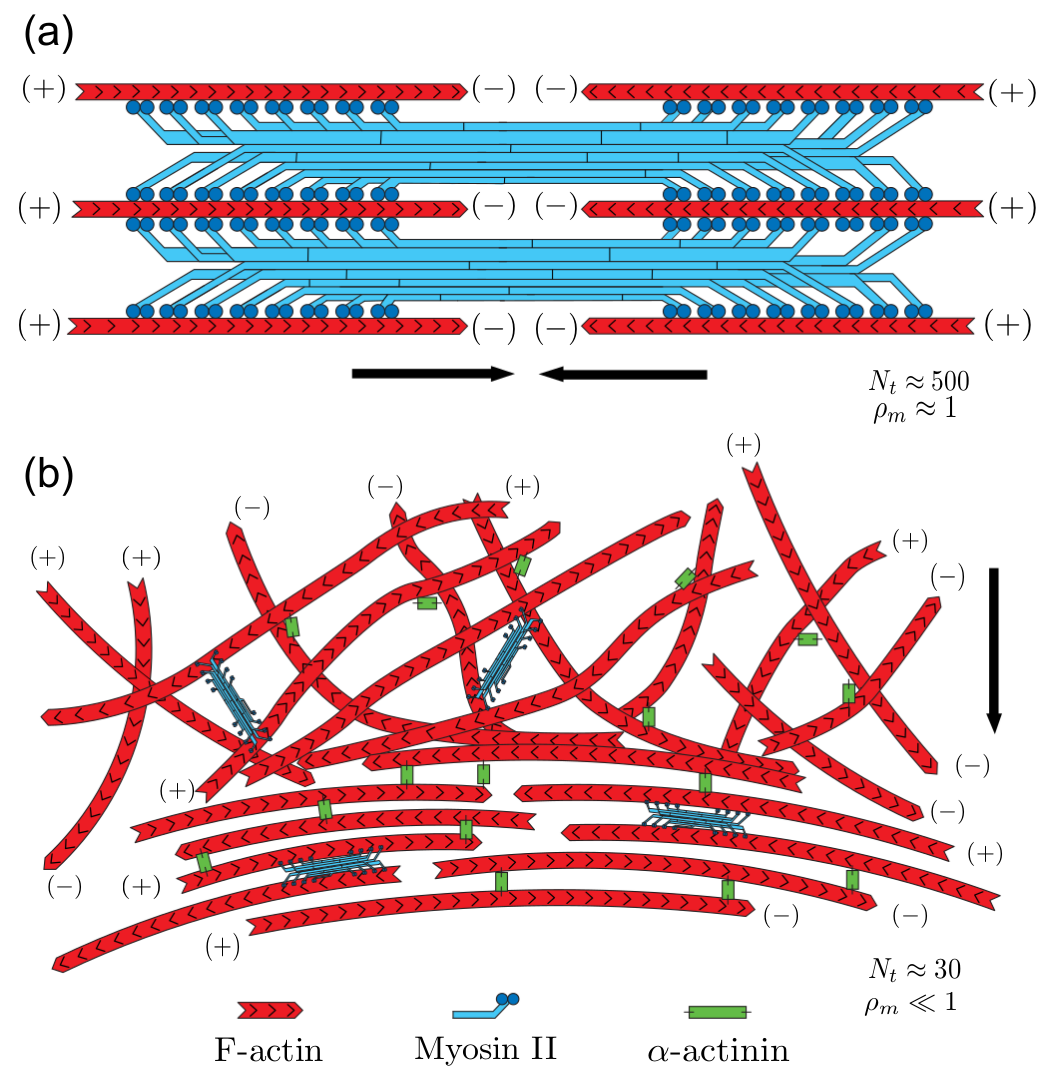}
  \caption{Actomyosin contractility mechanisms in a variety of cell types. (a) In a muscle sarcomere, actin filaments are aligned in bands of opposite polarity such that a bipolar myosin II filament can walk towards the actin filament's plus ends, generating maximal contractile force, shown as black arrows. The motor filaments also contain hundreds of heads, which are able to continuously generate force on the filaments and maintain attachment as they hydrolyze ATP to produce mechanical work. (b) In a non-muscle actomyosin network, filaments are distributed in a random geometric fashion throughout the cytoskeleton. The myosin II filaments in a non-muscle actomyosin system are also smaller (number of heads per side of bipolar filament $N_t\approx30$ \cite{Billington2013}) and highly transient (the duty ratio of bound to unbound states of the motor filament $\rho_m \ll 1$) compared to their muscle counterparts ($N_t \approx 500$, $\rho_m\approx1$), but can form disordered arrangements of locally anti-parallel actin filaments. Motor filaments of this nature are responsible for stress fiber formation via compression of a fragmented lamellipodial actin mesh.}
      \label{fig:1}
\end{figure}

Many eukaryotic cells fundamentally rely on the ability to reshape and reform their interior cytoskeletal polymer structure to produce directed motion \cite{Alberts2002}. This biological active matter, which constantly consumes and dissipates energy in its surrounding environment, exerts directional forces allowing the cell to dynamically respond to a variety of chemical or mechanical extracellular cues~\cite{Schmidt1998,Vogel2006,Luo2013}. An intriguing example of a cytoskeletal system that drives these cellular morphological changes is the cell's actomyosin network - a polymeric network of long, thin actin fibers which can be nucleated and rearranged via actin binding proteins and active, ATP-consuming molecular motors to produce cellular forces in a variety of structures, providing a dynamic scaffold for the cell body. The distinct ability of the combination of actin filaments, passive actin cross-linking proteins and myosin II molecular motors to produce contractile cellular force is of fundamental importance in many cell types and has been well-studied in muscle cells \cite{Hill1939, Duke1999, Vilfan2003}, where parallel arrangements of filaments in opposite polarities allows for a directed, inward pull of the cell's sarcomeric unit (Fig.~\ref{fig:1}). 

Contractility is, however, much less understood in motile non-muscle cells, where it frequently emerges from an isotropic filament network with little geometric or polar ordering, indicating some type of a spontaneous symmetry breaking process~\cite{Salbreux2012, Levayer2012}. These contractile networks are responsible for diverse micro-structural dynamics, including lamellar stress fiber formation of tens of bundled filaments (shown in Fig.~\ref{fig:1}(b)), cell rear retraction that is essential to locomotion, and tension maintenance in the thin sub-membrane cortical layer surrounding the cell~\cite{Blanchoin2014a}. Prior works on explaining the emergence of contractility in disordered actomyosin networks have pointed to a variety of effects, including the potential importance of filament buckling~\cite{Lenz2014,Lenz2012, Murrell2012, SoareseSilva2011}, actin filament treadmilling~\cite{Oelz2015, Popov2016a}, and the role of passive cross-linkers as force-transmitters between neighboring contractile clusters~\cite{Bendix2008, Kohler2012, Wang2012, Alvarado2013, Jung2015, Ennomani2016a, Popov2016a}. Other studies have also investigated the general non-equilibrium dynamics of such cytoskeletal arrangements, outside of the context of the emergence of contractility, and have described fluidized and glassy network behavior \cite{Wang2011, Popov2016a} as well as strain-stiffening of the active polymer gel \cite{Levine2009a}.

In light of both recent and older works, the fundamental physical intuition for the emergence of contractility still largely rests on the classical sarcomeric mechanism, where myosin mini-filament heads are bound to two locally anti-parallel actin filaments, allowing the energy-consuming unidirectional walking of those heads to generate a contractile shearing motion between the actin filaments (shown in Fig.~\ref{fig:1}(a)). Two questions naturally arise with regard to this picture: (1) How do initially isotropic actin networks in three dimensions give rise to locally anti-parallel contractile force dipoles, and (2) do passive cross-linkers play an important role only in percolating force chains at larger spatial scales, as previously suggested, or are they also important in generating contractility at the level of a single force dipole? We address the first question in the accompanying paper. 

In this work, we address the second question by analyzing a simple model of two anti-parallel actin filaments, connected at their furthest ends by springs to boundaries, allowing both cross-linkers and a myosin mini-filament to transiently bind and dissociate. Our analytical and numerical results, in qualitative agreement with recent works discussed below \cite{Walcott2010, Lansky2015}, indicate that passive cross-linkers play a crucial role already at the scale of a force dipole, giving rise to a funneled energy landscape, where stronger overlap of actin filaments corresponds to a larger number of bound cross-linkers, with those configurations having lower free energy compared with less contracted states. In addition to shedding light on this gradient of free energy due to cross-linker binding, we have also discovered a strongly biphasic behavior of contractility with respect to the strength of cross-linker binding free energy, where a rise of contractile forces is followed by a rapid decline due to kinetic arrest of intermediate configurations by strongly bound cross-linkers. Overall, we found that the thermodynamic drive to more complete overlap of actin filaments is significantly attenuated by trapping in a purely passive force dipole (pFD), generating contractile forces that are noticeably smaller than the ones generated by the unidirectional walking of myosin heads when in the presence of stiff springs counteracting the contraction. However, pure myosin action is also relatively ineffective at generating contractile forces because the bound motor filaments are highly transient, easily permitting recoiling slippage of the contracted element of two oppositely polar actin filaments. We have discovered that cross-linkers can help to overcome this slippage process via a ratcheting mechanism - by using dynamic cross-linkers to prevent recoiling of intermediate configurations without bound motors, greatly amplified contractile forces are produced. Hence, our work reveals strong synergy between passive cross-linker binding dynamics and active myosin processes in an active force dipole (aFD), which constitutes the main building block of actomyosin contractile network.

\section{Modeling a force dipole}

\begin{figure}
\includegraphics[width=3.5in]{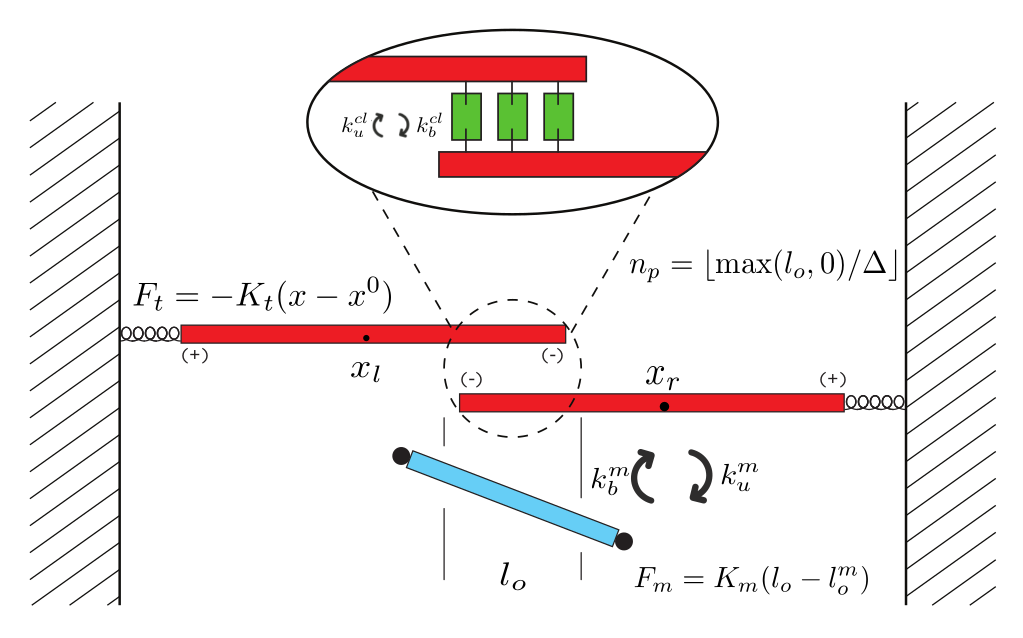}

\caption{One-dimensional "force dipole" actomyosin model schematic. Two filaments with plus ends facing outward and midpoints $x_r$ and $x_l$ are both connected to springs with stiffness $K_t$, initially with an equilibrium filament position of $x_r^0$ and $x_l^0$. When no cross-linkers are bound between the two filaments, they seperately undergo tethered Langevin motion. While the filament overlap $l_{o}$ is large enough, cross-linkers can transiently bind and unbind according to their kinetic rates as in Eq.~\ref{eq:kinetics}, which arrests the filaments if the number of cross-linkers bound is non-zero.The number of available binding sites $n_p$ varies with actin filament overlap. Motor filaments can (un)bind and walk stochastically on the pair of actin filaments, generating force via a time-varying filament overlap potential.}
\label{fig:model}
\end{figure}

To shed light on the thermodynamic and kinetic nature of a fundamental contractile actomyosin element interacting with passive cross-linkers, which serve as building blocks of more complex actomyosin networks in non-muscle cells, we constructed and carried out various simulations of a pair of rigid actin filaments in one dimension (Fig.~\ref{fig:model}), where cross-linker proteins dynamically bind and unbind to the overlapping region. We first considered a system where only cross-linker binding generates contractile forces in a passive force dipole (pFD), followed by simulations where a uni-directional motor was added to form an active force dipole (aFD), in which the motor filament can walk in the direction of the plus end of the actin filaments to generate additional mechanical force. The two actin filaments, denoted as the right and left filaments with respect to the simulated length, have length $L=2 \mu m$ and midpoints $x_r$ and $x_l$, with initial midpoint positions $x_r^0=1\mu m$ and $x_l^0=3\mu m$, respectively. An overlap length $l_{o}$ between the two filaments can then be defined as $l_{o} = (x_l + L / 2) - (x_r - L / 2)$. Each filament is connected to springs at their outward-facing plus ends such that the tension provided by the springs on each filament is $F_{t}^r = - K_t (x_r - x_{r}^0)$ and $F_{t}^l = - K_t (x_l - x_{l}^0)$. The diversity of the chemical states of the force dipole can be represented with the integer values, $(m,n)$,  with $m \in [0,1]$ and $n \in [0,\infty]$, which specify the number of (active) myosin II motor filaments and (passive) cross-linkers bound to the filament pair, respectively. 

When there are no motors or cross-linkers bound to the pair of filaments ($m=0$,$n=0$), they can separately undergo overdamped Langevin motion that includes forces provided by their respective tethers and a stochastic force, via $\eta \frac{dx_l}{dt} = F_{t}^l + F_{s} $, and $\eta \frac{dx_r}{dt}  = F_{t}^r + F_{s} $, where $F_s$ satisfies the fluctuation-dissipation theorem in one dimension. The instantaneous number of sites, $n_p$, available for binding in the overlapping region between the two filaments, is  obtained from $n_p = \lfloor \text{max}(l_{o},0) / \Delta \rfloor$, where $\Delta$ denotes the spacing between adjacent binding sites. We assume that the filaments move only in the absence of bound cross-linkers ($n=0$), otherwise, they are immobilized ($n\neq0$). We further expect that upon binding of a cross-linker, the tension in each respective actin filament tether is nearly instantaneously equilibrated such that $ |F_t| \equiv |F_t^r| = | F_t^l |$ while preserving the average tensile force experienced by the filament pair at the time-step before equilibration $|F_t| = \frac{1}{2}(|F_t^l| + |F_t^r|)$. This equilibration assumption is based on the separation of timescales between tether relaxation ($\approx ms$) and (un)binding dynamics ($\approx 0.1s$ in the fastest simulated case). The spatial positions of the filament pair remain stationary until complete cross-linker unbinding ($n=0$).

The filament pair can also experience an active motor contractile force, $F_m$, when a motor is bound ($m=1$), via a stochastically time-varying filament overlap potential, $F_m(t) = K_m (l_{o}(t) - l_{o}^{m}(t))$, where $l_{o}^{m}(t)$ indicates the filament overlap corresponding to an unstretched motor. Upon a new binding event, this motor parameter, which  implicitly represents the intrinsic length of the molecule, is initialized to $l_o(t)$. This defines the motor filament's equilibrium length in terms of the filament pair's configuration at the moment of motor's binding. Then, $l_{o}^m(t)$ can stochastically increase by a motor step size $d_s$ with average walking velocity $v_w$, where this velocity depends on the instantaneous value of $F_m$, as described in Appendix A. With this active force contribution, the aggregate force experienced by all cross-linkers bound between the filament pair can be written as $F_{cl} = |F_m - F_t|$. The simulation protocol iteratively performs Gillespie stochastic simulation for chemical dynamics (including both binding, unbinding and motor-stepping events) and switches to overdamped (mechanical) Langevin dynamics when the number of bound motors and cross-linkers becomes zero (i.e. $m$=0 and $n$=0). The latter dynamics is then evolved for $\tau_{r}$, which is the reaction time for the next re-binding event, estimated stochastically based on the same Gillespie algorithm. 

Next, we relate the stochastic (un)binding rates of cross-linkers on the filament pair to their thermodynamic binding energy, denoted as $\epsilon$. This binding energy can be phenomenologically related to the change in Helmholtz free energy $\Delta A$ by the following form, which contains enthalpic and entropic contributions~\cite{Hermans2014}:

\begin{equation}
\label{eq:bindingenergy}
\Delta A \approx - \epsilon + k_b T ln\Bigg(\frac{v_{m}}{v_b}\Bigg)
\end{equation}

\noindent where $v_{m} = V / N$, $V$ is the system volume, $N$ is the number of cross-linkers in solution. $v_{b}$ can be thought of as the binding site volume of the cross-linker, which at most physiological concentrations is much smaller than $v_{m}$. Intuitively from this equation, one can imagine system entropy is lowered from binding as $k_b T ln(\frac{v_{m}}{v_b})$, but is counter-acted by a gain in energy due to the favorability of cross-linker binding with energy $\epsilon$. 

We assume upon cross-linker binding $P \Delta V \approx 0$ such that we can equate the change in Helmholtz free energy $\Delta A$ to the change in Gibbs free energy $\Delta G$. Combining Eq.~\ref{eq:bindingenergy} with this approximation, and using the classic relation for the dissociation of a molecule from a binding site $K_D = e^{\Delta G / k_b T}$, one obtains an approximate expression for the dissociation constant $K_D$ of cross-linkers in this system in terms of their binding energy as

\begin{equation}
\label{eq:kinetics}
K_D \approx \frac{v_{m}}{v_b} e^{-\epsilon / k_b T}.
\end{equation}

\noindent $K_D$ can then be related to the stochastic cross-linker reaction constants by $K_D = k_{u}^{cl} / k_{b}^{cl}$. 

Non-muscle myosin II molecules \textit{in vivo} are assembled into bipolar mini-filaments, with patches of motor heads radiating outward \cite{Pollard1982} which then can (un)bind to neighboring filament segments. Hence, we must consider motor filament kinetics in our model as a coarse-grained version of a detailed stochastic process including $N_t$ transient motor heads per side of the mini-filament, coexisting in a connected bipolar structure. The number of motor heads per side of a mini-filament is typically $N_t \approx 30$ (in total $\approx 60$ heads per mini-filament) \cite{Billington2013}, so it is reasonable to assume that $<10$ of these heads per side could be bound to a single actin filament simultaneously - we choose $N_t=10$ for our simulations, which produces a mean unbinding time of $\bar \tau_u^{m} \approx 1/s$ in absence of mechanosensitivity. Since tension is released upon unbinding of either side of the bipolar filament, we use this value as the effective unbinding rate of our coarse-grained description. Therefore, even with multiple motor filament heads possibly attached to a single actin filament, the stochastic dynamics of these heads creates a highly transient system of tension build and release between the filament pair. We leave the description of our mechanochemical model for these myosin II filaments, which is used in our aFD simulations, to Appendix A.

We include catch and slip bonds for cross-linkers as well as motor filaments. Although the aim in this study is not to probe the effects of mechanochemical feedback between these proteins, we find it necessary to include key mechanochemical relationships for full model realism. For individual cross-linker bound to the filament pair, we employ a typical ``slip" bond characterized by a decreasing bound lifetime with applied load \cite{Ferrer2008} as $k_{u,eff}^{cl} = k_{u}^{cl} \text{exp}(F_{cl} x_{cl} / n_p k_b T)$ where $x_{cl}$ is a characteristic unbinding distance, and $k_{u}^{cl}$ is the zero-force unbinding rate. We only consider pulling forces such that $F > 0$. In the case of a non-muscle myosin II motor filament, where individual motor heads have been shown to have a ``catch"-bond characterized by an increased bound lifetime with applied load \cite{Guo2006, Kovacs2003, Kovacs2007}, we describe the (un)binding and walking kinetics of an ensemble of motor heads with a simple set of parameters to capture the essential aspects of motor filament mechanosensitivity - a binding lifetime that increases exponentially with applied load $F$ and the number of motor heads in the filament $N_t$, motivated by the results of \cite{Erdmann2013}, and a walking velocity that is one of the Hill form \cite{Hill1939}. In Appendix A, we describe the mechanochemical model of the myosin II filament used as well as all parameter values used for the overall model.

\section{\label{sec:level2}Passive force dipoles}

\begin{figure}
\includegraphics[width=3.2in]{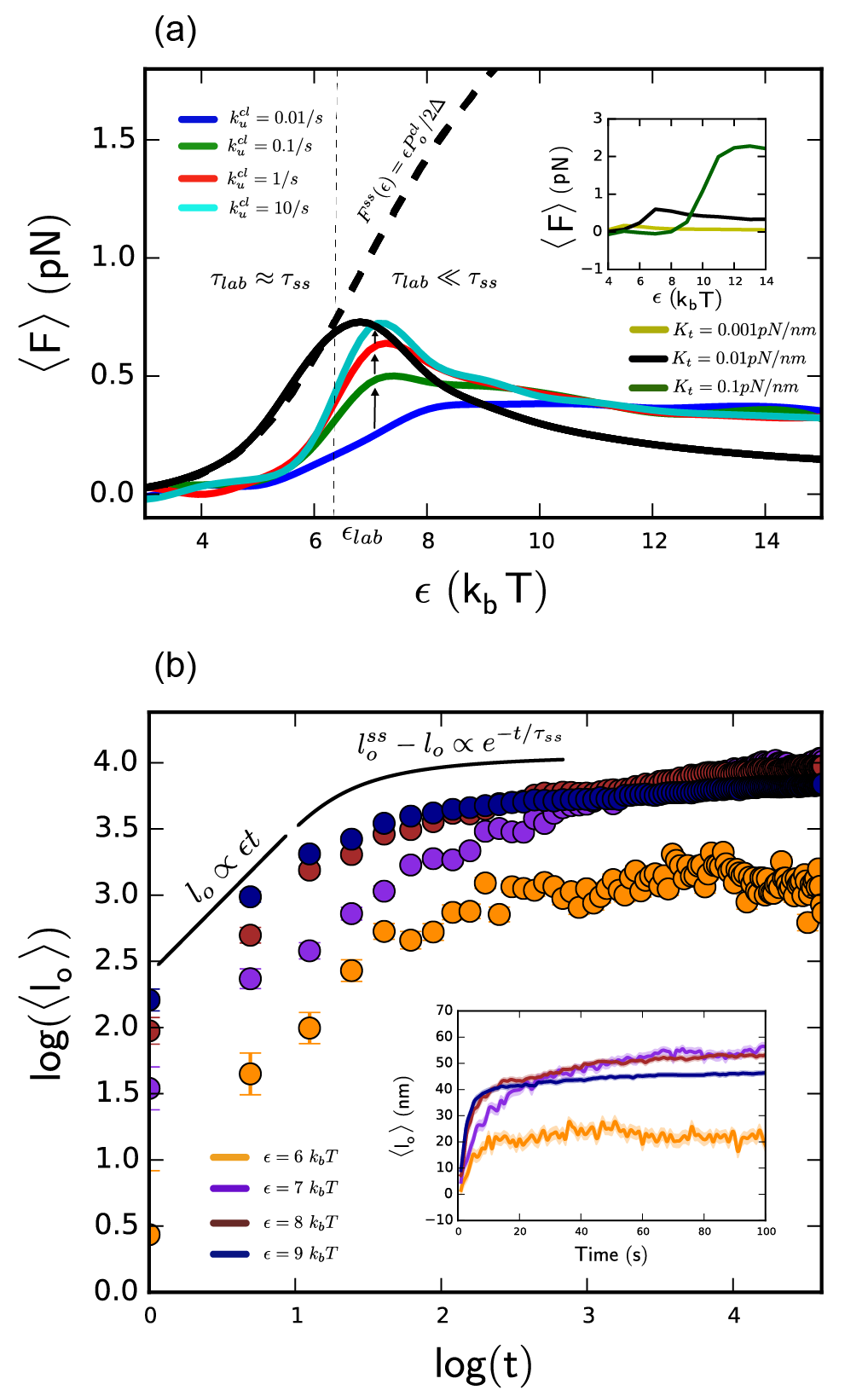}
    \caption{Cross-linker driven contraction of the passive force dipole. (a) Measurement of trajectory averaged contractile force on $\tau_{lab}$ for $K_t=0.01pN/nm$ tethers, denoted as $F(\epsilon)$ shows non-linear behavior in terms of the cross-link energy $\epsilon$ as chemical input with $k_u^{cl} = 10/s$. The analytically predicted $l_o^{ss}$ is shown as the black dotted curve, with the numerical solution to Eq~\ref{eq:effcrosslinkermotionavg} as solid black - these solutions diverge at $\epsilon_{lab} \approx 6 k_b T$. This corresponds to the transition $\tau_{lab} \ll \tau_{ss}$ caused by increasing binding strength of cross-links. Increasing speed of unbinding shifts curves to the predicted thermodynamic limit. Inset displays stochastic trajectories for the same cross-linker binding energies. (b) Power-law analysis of the trajectory-averaged motion of the filament pair shows a linear force-generating regime $\l_o \propto \epsilon t$ followed by a exponential relaxation to steady state $l_o^{ss} - l_o \propto e^{-t / \tau_{ss}}$ with $\tau_{ss} \propto (1+\text{exp}(\epsilon/k_bT))^{l_o/\Delta}$. Trajectories above $\epsilon_{lab}$ are far from equilibrium as seen in their steep decay to arrest.}
    \label{fig:nomotor}
\end{figure}

In the absence of motor filaments, cross-linkers can create a substantial amount of contractile force in the anti-parallel filament pair against the elastic tethers simply by rectifying Brownian fluctuations which increase filament overlap. We denote this as a passive force dipole (pFD) since it does not contain active fluctuations. Contraction of the pFD can be quantified in terms of a thermodynamic parameter $\epsilon$ and kinetic parameter $k_u^{cl}$ using their previously defined relation. Figure~\ref{fig:nomotor}(a) shows the average contractile force generated in 250 simulations for 200 $s$ - we denote this time as $\tau_{lab}$ and average trajectory observables $\langle \cdot \rangle$ over this time. We observe a sharply biphasic dependence of generated overlap on varying cross-linker binding energies. At low $\epsilon$, cross-linkers cannot sustain significant levels of tension between the filaments due to a low occupancy of the available binding sites, while at high $\epsilon$, significant kinetic arrest occurs due to quick saturation of the binding sites, significantly hindering filament motion. While cross-linkers are able to generate overlap between the filament pair when restoring forces are minimal, at most binding energies simulated, they are ineffective against stiffer springs counteracting the contraction of the pair ($K_t=0.1pN/nm$). This is due to slippage of the filament pair when cross-linker dissociation occurs, releasing contractile tension between them. Contraction can only be generated in this case if the binding affinity of the cross-linkers is greater than $10k_bT$, which induces complete arrest of the filament pair upon a single site occupancy.

Essential to the characteristic force production of the pFD is not only the overall cross-linker affinity $\epsilon$, which controls the probability of occupancy of the $n_p$ sites between the filament pair, but also, crucially, the kinetic rates of binding and dissociation processes that determine how quickly initially non-equilibrium dynamics approaches the steady state. The former arise purely from statistical mechanics principles, as we show below. Accelerating cross-linker binding kinetics in this regard (i.e. simultaneously increasing $k_b^{cl}$ and $k_u^{cl}$ while maintaining binding energy as in Eq.~\ref{eq:kinetics}) shifts resulting force curves upward until saturation is reached for $k_u^{cl} =$ $10$ $s^{-1}$. This can be thought of as approaching the thermodynamic limit of force production of the contractile element as $k_u^{cl} \rightarrow \infty$. 

Important to this non-linear behavior is the mean-first passage time of unbinding of $n$ cross-linkers from $n_p$ cross-link binding sites on the filament pair, $\bar \tau_{u}^{cl}$, since this mean behavior defines the speed of inter-filament motion across all possible overlaps, and thus the timescale of relaxation $\tau_{ss}$ of the contractile element. In our stochastic representation, this is the mean time of transitioning from the $n=1$ to $n=0$ state. Surprisingly, this mean-passage-time problem is similar to the stochastic dynamics of an ensemble of myosin II motor heads becoming completely unbound from an actin filament (see Appendix A, Eq.~\ref{eq:off0}). Recasting this equation for cross-linker dynamics by replacing the number of motor heads $N_t$ with the number of possible binding sites $n_p$, as well as cross-linker kinetics of (un)binding, we obtain an expression for the mean-passage-time of complete cross-linker unbinding:

\begin{equation}
\label{eq:off0cl}
\bar \tau_u^{cl} \approx \frac{1}{k_{b}^{cl} n_p} \Bigg[\Bigg(1 + \frac{k_b^{cl} }{k_u^{cl}}\Bigg)^{n_p} - 1\Bigg].
\end{equation}

\noindent On the other hand, the mean binding time (i.e. transition time from $n=0$ to $n=1$) is simply $\bar \tau_b^{cl} = k_b^{cl} n_p$. We note that both passage times are  fundamentally dependent on the filament overlap by the definition of $n_p$ given above. Predictions of Eq.~\ref{eq:off0cl} show excellent agreement with simulated first-passage times of cross-linkers from $n=1$ to $n=0$ at various binding site availabilities (data not shown).

%\subsection{An equation of motion describing the contraction process}

\begin{figure*}
\includegraphics[width=7.0in]{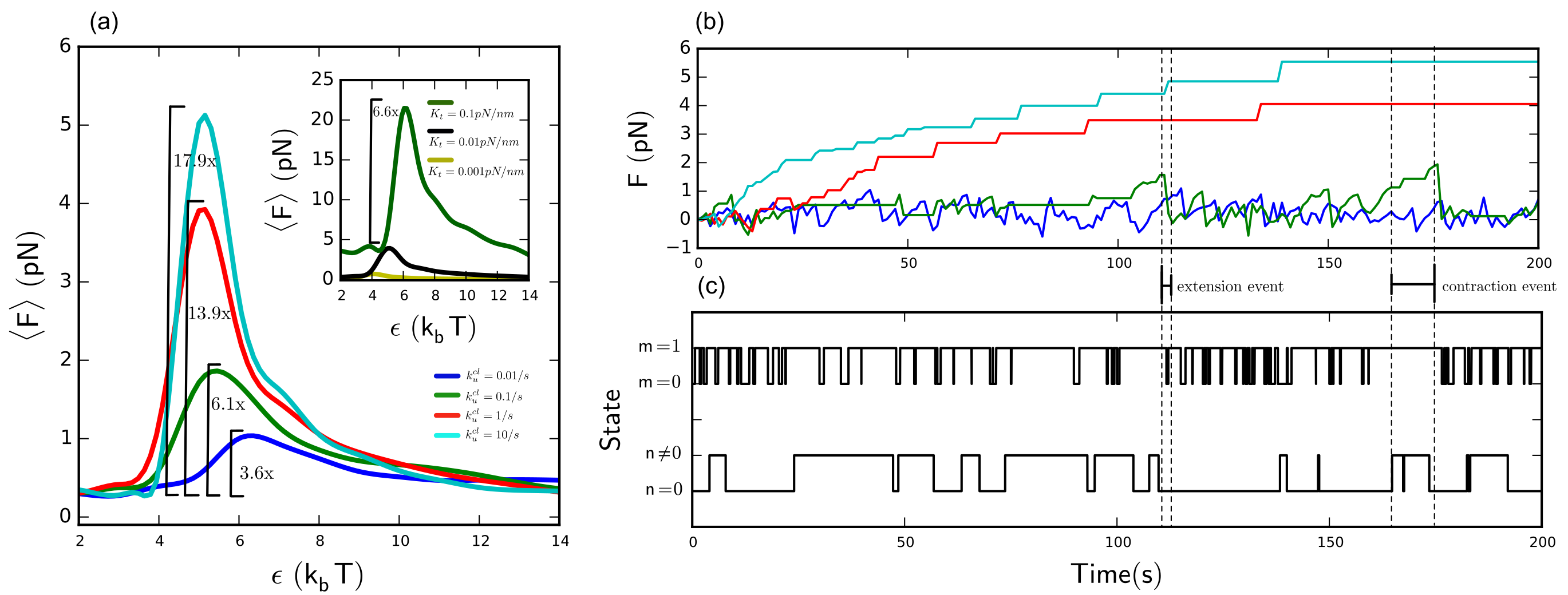}
    \caption{Actomyosin ratcheting due to transient cross-linking in the active force dipole. (a) $F(\epsilon)$ shows similar biphasic behavior but is amplified greatly due to the presence of the stochastic motor against a $0.01pN/nm$ tether. Amplifications from the transient motor force are shown (transient motor force $\approx0.3pN$). Inset shows the mechanosensitive response of the motor filament for increasing external stiffness with $k_u^{cl} = 1/s$. (b) Trajectories of the filament pairs at $\epsilon_{lab} \approx 5 k_b T$ show step-like approach to a steady state. Increasing kinetic rates maximize overlap generated on $\tau_{lab}$. (c) The corresponding state trajectory of the simulation with $k_u^{cl}=0.1/s$, which can determine contraction and extension events in (b). $m=0,1$ represents the unbound and bound states of the motor, respectively. $n=0$ and $n \neq 0$ represent the cross-linker state.}
    \label{fig:udmotor}
\end{figure*} 

With this, we can construct an equation of motion describing the observed stochastic contraction process in terms of the cross-linker's mean-field energetic contribution to the filament pair and the resulting elastic restoring forces of the filament pair. First, in absence of stochastic effects, a thermodynamic driving overlap force $\epsilon P_{o}^{cl} / \Delta$ is expected to result when the filament pair increases their overlap by a distance $\Delta$, with a probability of occupancy of the binding site $P_{o}^{cl} = k_{b}^{cl} /  (k_{b}^{cl} + k_{u}^{cl}$). Since the motion of the filament pair is hindered if $n\neq0$, an effective mean-field equation of motion neglecting stochastic force can be written as:

\begin{eqnarray}
\label{eq:effcrosslinkermotion}
2 \eta \frac{dl_o}{dt} = \bigg(\frac{\epsilon}{\Delta}P_{o}^{cl} - 2 K_t l_o\bigg) (1 - \delta_n),  \text{ where} \\
\delta_n = 
\begin{dcases}
		0, & \text{if } n = 0. \\
		1, & \text{if } n \neq 0.
\end{dcases}
\end{eqnarray}

\noindent We have defined $\delta_n$ such that motion is allowed when there are no cross-linkers bound to the filament pair, and $\eta = 10^{-3} pN s / nm$ is the viscous damping constant of an individual actin filament. If we average this equation over many intervals of $n=0$ and $n\neq0$ states, which is valid under the assumption that many intervals occur in motion to a new cross-link position $l_o \pm \Delta$, i.e. $\eta \Delta^2 / \epsilon \gg \bar \tau_b^{cl} + \bar \tau_u^{cl}$, the $\langle (1 - \delta_n) \rangle_{ia}$ term can be separated from the average and evaluated explicitly as the probability of the $n=0$ state during simulation. This assumption bounds the kinetic rates considered to $\bar \tau_u^{cl} \lesssim 1 s$. We express this probability in terms of the cross-linker binding energy as $P(n=0 | l_o) = \bar \tau_b^{cl} / (\bar \tau_b^{cl} + \bar \tau_u^{cl}) =  (1+v\exp(\epsilon))^{-l_o/\Delta}$, which bears similar resemblance in its form to the classic statistical mechanics problem of Langmuir absorption. We have defined $v = v_{m} / v_b$. This then gives the final equation of motion, where we change variables $\tilde l_o$ to indicate the interval-averaged overlap:

\begin{equation}
\label{eq:effcrosslinkermotionavg}
2 \eta \frac{d \tilde l_o}{dt} =  \bigg(\frac{\epsilon}{\Delta} P_{o}^{cl} - 2 K_t \tilde l_o\bigg) (1+v\exp(\epsilon / k_b T))^{-\tilde l_o/\Delta}.
\end{equation}

\noindent This equation predicts a near-linear contractile regime in the short-time limit with a constant velocity: $l_o \propto v_{o}^{cl} t$. This is followed by an exponentially decaying relaxation to steady state $l_o^{ss} - l_o \propto e^{-t / \tau_{ss}}$ (details of the asymptotic analysis are given in Appendix B). In general, $v_{o}^{cl}$ is directly proportional to the strength of cross-linker binding in absence of non-equilibrium effects, and $\tau_{ss}$ is inversely proportional to the occupancy of the available binding sites. This is observed in our simulations with predicted $v_{o}^{cl} \propto \epsilon$ and $\tau_{ss} \propto (1+\text{exp}(\epsilon/k_bT))^{l_o/\Delta}$ asymptotic behaviors, as shown in Fig~\ref{fig:nomotor}(b). At infinitely long timescales, the equation of motion predicts the steady state overlap,

\begin{equation}
\label{eq:eloss}
l_o^{ss}(\epsilon) = \frac{\epsilon P_o^{cl}}{2 \Delta K_t},
\end{equation}

\noindent which is shown in Fig~\ref{fig:nomotor}(a) as the upper bound of the finite-time solutions which contain kinetic arrest. This steady-state prediction diverges from measurements during the laboratory timescale (i.e. $\tau_{ss} \gg \tau_{lab}$) when $\epsilon$ passes a threshold of $6k_bT$, which we denote as $\epsilon_{lab}$. Therefore, our results indicate the sharp onset of a glass-like behavior which produces a transition to far-from-equilibrium filament states. This greatly limits force production of the filament pair when cross-linker affinity becomes greater than $\epsilon_{lab}$. 

We have seen that a passive force dipole can produce low $pN$ scale forces if the cross-links are chemically favored to bind to the filament pair from the surrounding solution, forming a funneled free-energy landscape in $l_o$. Despite the manifest non-equilibrium nature of the contraction process, the stochastic trajectories of the dynamically cross-linked two-filament element could be well understood by treating cross-linkers as exerting a mean-field mechanical driving force of thermodynamic origin.  Furthermore, our analysis predicts that this driving force acts to not only favor but also eventually arrest filament overlap depending on the interaction strength between cross-linkers and actin and the resulting speeds of cross-linker binding and unbinding.  

\section{Active force dipoles}

We now also investigate the positive feedback mechanism which produces an apparent force amplification of the cross-linked filament pair when in the presence of active fluctuations. As highly transient myosin II motor filaments are added (the mechanochemical model for motor filaments is outlined in Appendix A) to form an active force dipole (aFD) in absence of cross-linkers, the frequent unbinding of the motor filament causes a continual build and release of contractile tension over the simulation duration, resulting in minimal force generation ($\approx0.3pN$ against $K_t=0.01pN/nm$ springs). However, when cross-linkers are present, 3- to 17-fold amplifications are observed compared to the original forces generated in absence of the motor filament (Fig~\ref{fig:udmotor}) at an $\epsilon_{lab} \approx 5 k_b T$. This is slightly shifted from the pFD case due to cross-linker mechanochemical effects (data not shown). We also observe an overall sharp increase in transient motor filament force production when increasing stiffness of external springs, consistent with the myosin II's catch bond nature \cite{Erdmann2013, Stam2015}. At higher external stiffness $K_t = 0.1pN/nm$, motor filaments alone are more effective than cross-linkers at generating contractile forces for a range of binding affinities ($\epsilon\approx0-10k_bT$) but are still minimally contractile compared to when both passive and active elements are present, which produces $22 pN$ of force at peak binding affinity, approaching the motor filament's stall force $F_s =24pN$. The collective behavior of the one-dimensional actomyosin-cross-linker system is shown in Fig.~\ref{fig:udmotor}(b)-(c) - a ``ratcheting" behavior in which the motor, although becoming (un)bound frequently, can steadily produce force over the entire simulation interval, helped by cross-linkers transiently stabilizing filament overlaps. The power-law behavior of the active element displays similar characteristics compared to when motor filaments are absent in the passive element (see details in Appendix B).

\begin{figure}
\includegraphics[width=3.2in]{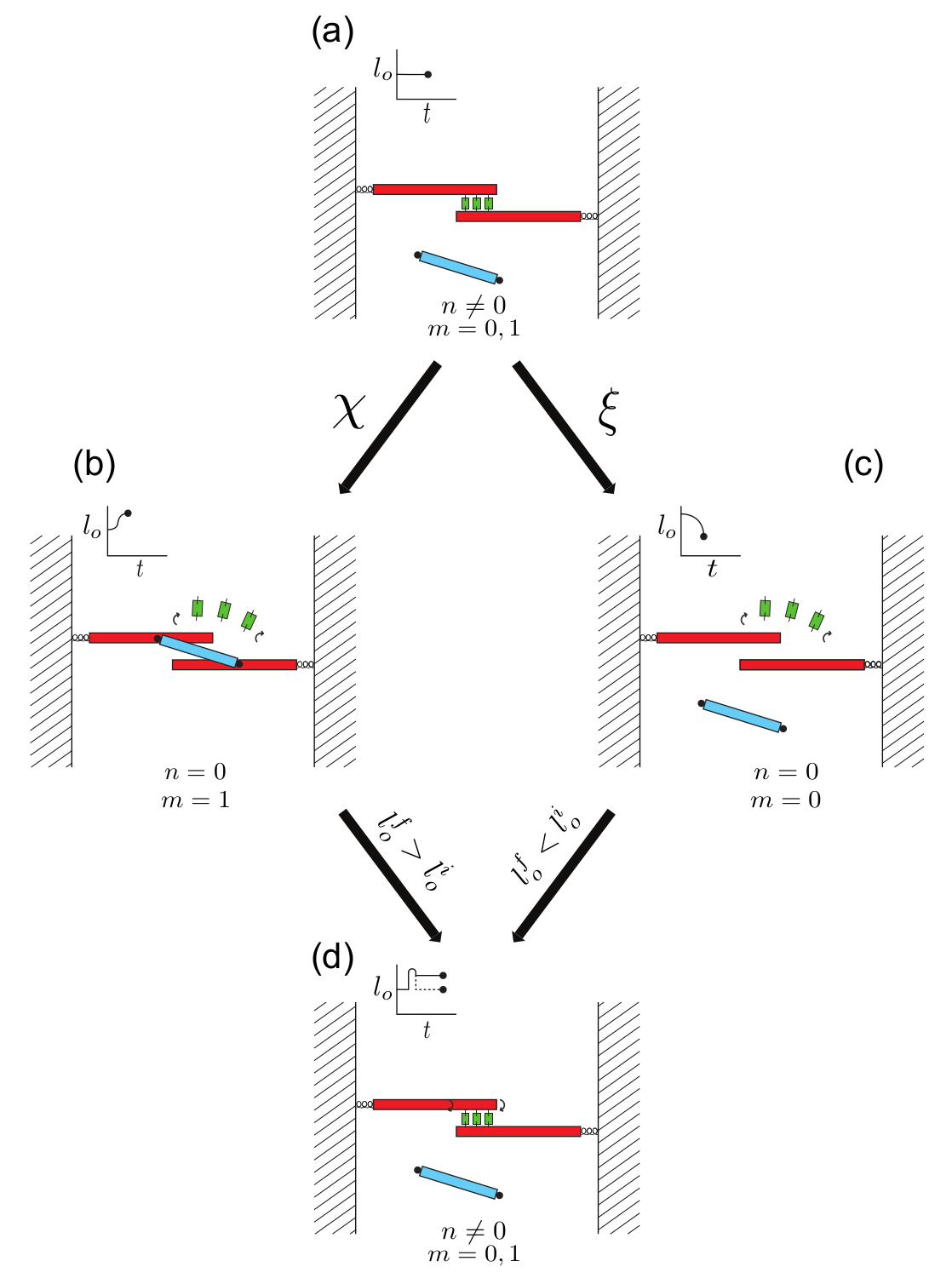}
    \caption{The proposed ratcheting process in an active force dipole. (a) The process starts at a stable configuration $n=0$. The motor can be either bound or unbound ($m=0$ or +1). (b) If $m=+1$, a contraction event occurs, contracting the filaments to a new overlap $l_o^{f} > l_o^{i}$. (c) If $m=0$, an extension event occurs in which the filaments lose overlap such that $l_o^{f} < l_o^{i}$. (d) Cross-linkers re-bind ($n\neq0$), stabilizing the $l_o^{f}$ achieved in (a)-(b) - in a contracted or extended configuration. This process repeats as pairwise overlap is created between the anti-parallel pair, generating contractile force.} 
    \label{fig:ratchet}
\end{figure}

\begin{figure}
\includegraphics[width=3.4in]{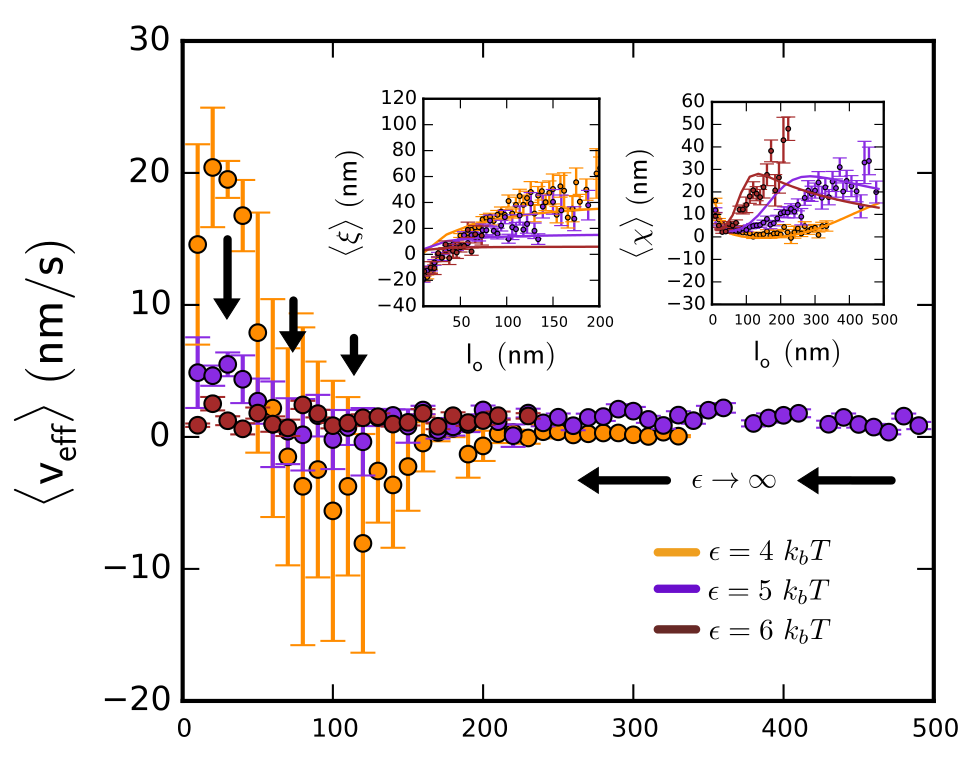}
    \caption{Effective inter-filament velocity profile in the active force dipole. (a) The presence of cross-linkers can trap intermediate filament configurations in the the aFD with $k_u^{cl} = 10/s$ against $K_t=0.01pN/nm$ tethers. The effective velocity is of the element is calculated as $v_{eff} (l_o) =  \rchi (l_o) \cdot \omega_\rchi(l_o) - \xi (l_o) \cdot \omega_\xi(l_o)$. This amplifies the resulting forces produced by the element and is apparent in the effective contractile velocity of the filament pair. Analytic approximations for the expected value of these events are showed as filled lines. }
    \label{fig:ratchetdata}
\end{figure}

While the complex stochastic and mechanochemical nature of the myosin II filament, in general, cannot be represented by an effective motor force term in Eq.~\ref{eq:effcrosslinkermotionavg}, we can quantify the efficiency of this aFD ratcheting (Fig.~\ref{fig:ratchet}) over the range of filament overlaps and kinetic parameters of the cross-linkers, to observe the specific effects of cross-linker affinity on motor dynamics. To do this we define two classes of stochastic events, both quantifying an overlap length change of the aFD upon a ratcheting cycle: a contraction event $\rchi (l_o)$ in which ($n=0$, $m=1$) and a extension event $\xi(l_o)$ in which ($n=0$, $m=0$), both of which end when $n\neq0$. The frequency and magnitude of these events will vary significantly based on the (un)binding dynamics of the cross-linkers which sharply regulate motion of the anti-parallel pair as previously described. By collecting this data from simulations, we can construct an effective inter-filament velocity profile of the aFD: $v_{eff} (l_o) =  \rchi (l_o) \cdot \omega_\rchi(l_o) - \xi (l_o) \cdot \omega_\xi(l_o)$, where $\omega_\rchi(l_o)$ and $\omega_\xi(l_o)$ represent the frequencies of the respective stochastic events from simulations. This can be seen as similar to a force-velocity relation since in our one-dimensional framework, the overall restoring force experienced by the aFD is directly proportional to the formed overlap between the anti-parallel filaments.

With this ratcheting data as shown in Figure~\ref{fig:ratchetdata}, it is clear that cross-linker dynamics can provide a bias for contraction in the presence of myosin II as seen in the significant shifting of the force-velocity relation of the element. Beginning with $\epsilon < \epsilon_{lab}$, large contraction events exist but not for substantial overlap, resulting from the inability of the cross-links to quickly re-stabilize contractile configurations and releasing tension. At $\epsilon \approx \epsilon_{lab}$, the contraction events are robust across all configurations due to sufficient pairwise trapping by the cross-links, producing a near-constant effective velocity. In the regime $\epsilon>\epsilon_{lab}$, contraction events are efficient and extension events minimal, but arrest increases significantly, hindering filament motion at large overlaps. In this view, increasing the speed of (un)binding events, as observed previously in the no-motor filament case, will alleviate kinetic frustration between the filament pair which allows maximal overlap generation (data not shown). This is due to inter-filament motion becoming more fluid-like in the presence of highly transient cross-links. We conclude that passive cross-links can shift contraction favorability of the actomyosin element by promoting stabilization at large filament overlaps where restoring forces are significant, as indicated by the robustness the force-velocity relation of the element at $\epsilon_{lab}$. This averaged behavior can be reproduced analytically by probabilistically averaging the possible kinetic events of contraction and extension in the element - this is briefly derived in Appendix C. 

We have shown that the energy landscape induced by transient cross-linking in a passive force dipole is directly observed in the non-equilibrium active force dipole (aFD) when transient motor filaments are also present. This results in a force amplification of the element via a ratcheting of the motor filament on the funneled energy landscape induced by the presence of passive cross-linkers. This amplification produces many $pN$ of force against elastic tethering, and allows the transient motor filament to reach its maximum stall force.

\section{Discussion}

In this paper, we have investigated the ability of cross-linked actin force dipole to contract by rectifying thermal and active motions in a simple stochastic model. The contraction of these elements is essential in understanding the macroscopic dynamics of cross-linked actin gels undergoing myosin II rearrangement in non-muscle systems, but differs from the classic sarcomeric contractility mechanisms that neglect passive cross-linkers as potential force producers. We found that the free energy landscape induced by the presence of cross-linkers can create noticeable contractile force in the filament pair, however, in a narrow window of binding affinities. To better understand the underlying kinetic restrictions, we developed a mean-field model, namely, a differential equation of motion describing the dynamics of intermittently arrested inter-filament states, where the cross-linker binding energy and the unbinding rate were the main varied thermodynamic parameters.  When cross-linker binding affinity is high, the resulting steeper free energy gradient with regard to increasing filament overlap (i.e. energy funnel) generates significant thermodynamic contractile forces, however, we also predict quick onset of a glass-like regime that will severely hinder the force production of the element. Our far-from-equilibrium treatment crucially alters the previously proposed linear dependence of the contractile force of filaments due to cross-linker binding \cite{Walcott2010, Lansky2015}, $F \propto \epsilon$, suggesting, instead, a sharply peaked force--binding-energy distribution, that exponentially decays at large binding energies. 

While passive cross-linkers in absence of myosin II lead to force production within the pair of filaments (passive force dipole), these forces in general are smaller than forces generated by myosin II motors. The latter by themselves also become rather inefficient, however, in the presence of significant tension at the boundary of the filament pair (active force dipole), due to overlap slippage caused by restoring forces when myosin II transiently dissociates, which is frequent because of their low processivity. Hence, in this work, we have discovered that simultaneous combination of all three components, namely actin filaments, cross-linkers and myosin II, can produce a highly contractile active force dipole due to a stochastic amplification mechanism elaborated above. Physically, this amplification is due to cross-linkers trapping overlapped states as transient motors (un)bind frequently, producing a ratcheting behavior which rescues tension release when motors become unbound. In the presence of these active fluctuations, the funneled landscape induced by passive cross-linking results in a further shift towards in contractile configurations evident in the force-velocity relation of the element. This strong amplification, which produces many $pN$ of force, is plausible since non-muscle myosin II motor filaments are highly transient \cite{Billington2013} and cross-linker kinetics are of comparable timescales \cite{Murrell2015} to motor head stepping and (un)binding. Two classes of experiments will aid in testing our theory: (1) single-molecule force measurements and distributions of residence times of small ensembles of non-muscle myosin II for more accurate models of the true minifilament dynamics, and (2) \text{in vitro} microrheology of cross-linked actin gels with those minifilaments to measure the stochastic fluctuations at a scale of filament pairs, building on the previous works \cite{Mizuno2007, Falzone2015}, however, with both cross-linkers and myosin II motors included in the same network. These results will deepen our understanding of the thermodynamically driven aspect of contraction in actomyosin cross-linked networks due to passive cross-linker binding (which we propose to denote as the ``contraction funnel''), as well as crucially important non-equilibrium effects due to the stochastic mechano-chemical dynamics of the proteins. 

In summary, we have shown that robust contraction can occur in the fundamental non-muscle actomyosin building block due to a mechanism arising from the energetic favoring of increased passive cross-linking, independent of mechanical network symmetry breaking (i.e. buckling of filaments as previously shown \cite{Murrell2012, SoareseSilva2011}). Overall, we envision a disordered actomyosin network to be comprised of highly contractile pFD's and aFD's, where the relative importance of these elements as well as mechanical asymmetry is largely determined by boundary stresses and overall actin network architectures. It will be interesting in future works to consider contractile processes in the context of both contractile mechanisms, and in more realistic three-dimensional environments. 

%Furthermore, passive cross-linkers and active myosin II act highly synergistically, via a ratcheting mechanism, leading to highly efficient contractile elements. Thus, our results suggest that contractile force generation in randomly oriented actomyosin arrays could occur if force dipoles exist in the network, and the extension of other filament pairs may be attenuated by the induced contractile landscape. This is relevant when studying the cell's lamellar region, where taut bundles spontaneously emerge from isotropic actin networks with myosin II and cross-linkers \cite{Verkhovsky1995}. 

\section*{Acknowledgements}

We thank Alex Mogilner for helpful discussions during the initial phase of this project. We are also grateful to Aravind Chandrasekaran, Cal Floyd, and Hao Wu for their review of the manuscript. All simulations were performed on the Deepthought2 Supercomputer at the University of Maryland. This work was funded by National Science Foundation Grant CHE-1363081.

\section*{\label{sec:levelAC}Appendix A - Mechanochemical parameter considerations for the simulated proteins}

The parameters chosen for our simulations are given in Table~\ref{table:params}, which includes experimentally measured mechanical stiffnesses as well as chemical kinetics of (un)binding for the proteins considered. Careful consideration must be made when choosing kinetic parameters for both cross-linkers and myosin II motor filaments, since these rate constants determine the stochastic dynamics of the force dipole. Considering cross-linkers \textit{in vivo}, the abundant cross-linker $\alpha$-actinin displays a bound lifetime of $~2.5s$ with free energy of binding estimated at $\Delta G \approx 3 k_bT$, as addressed in a review by Murrel et al \cite{Murrell2015}. Other abundant cytosolic cross-linkers, including fascin ($\Delta G \approx 15 k_b T$) and filamin ($\Delta G \approx 10 k_b T$), display a lifetime of $\approx 1s$ \cite{Murrell2015, Sun2010}. Thus, we choose to vary the binding energy control variable $\epsilon$ from 0 to 15 $k_bT$. We also use kinetic (un)binding rates for cross-linkers in the aforementioned physiological range while also testing slower kinetics: $k_u^{cl}$ is varied from $0.01/s$ to $10/s$.

\begin{table*}
\setlength\extrarowheight{3pt}
\centering
{\tablinesep=20ex\tabcolsep=14pt
\begin{tabular}{ | c | l | l |   } 
 \hline
\textbf{Parameter} &\textbf{Description} & \textbf{Value}  \\
 \hline
 \hline
 $k_bT$& Thermal energy & $4.1 \  pN nm$ \\
 \hline
 $x_{l,r}^0$ & Initial left (\textit{l}) and right (\textit{r}) actin filament midpoints & $1 \ \mu m, 3 \ \mu m$ \\  
 \hline
  $L$ & Length of actin filament & $2 \ \mu m$  \\
  \hline
  $K_t$ & Boundary tether stiffness & $0.001 - 1 \ pN/nm$ \\
  \hline
  $\eta$ & Viscous damping constant & $10^{-3} \ pN s / nm$ \\
  \hline
  $F_s$& Stochastic force experienced by actin filaments & - \\
  \hline
  $l_o$& Actin filament pairwise overlap & \textit{Observable}  \\
  \hline
  $F_t^{l,r}$& Tether forces experienced by left (\textit{l}) and right (\textit{r}) actin filaments & \textit{Observable} \\
  \hline
    $F_{cl}$& Force experienced by bound cross-linkers & - \\
  \hline
  $F_m$ & Force experienced by motor filament & - \\
  \hline
   $x_{l,r}$ & Instantaneous left (\textit{l}) and right (\textit{r}) actin filament midpoint & - \\
  \hline
  $l_o^m$& Motor-preferred actin filament overlap & - \\
  \hline
  $l_o^{ss}$& Steady-state overlap in pFD & -  \\
  \hline
  $N_t$ & Number of single motor heads per side of motor filament&$10$  $ ^\textbf{ a}$\\
\hline
  $m, n$ & Number of bound motor filaments (\textit{m}) and cross-linkers (\textit{n}) &  -  \\
  \hline
  $n_p$ & Number of possible cross-link binding sites & -  \\
  \hline
  $\epsilon$ & Cross-linker binding energy & $0-15 \ k_b T$ \\
  \hline
  $v_{m}$ & Effective volume of cross-linker in solution& $1 \cdot10^{-3}$ $ \mu m^3$  $ ^\textbf{ b}$ \\
\hline
   $v_{b}$ & Approximate bound volume of cross-linker & $3 \cdot10^{-6}$ $ \mu m^3$  $ ^\textbf{ c}$\\
\hline
 $\Delta$ & Distance between cross-link binding sites &$10 \ nm$  \\
\hline
   $k_{b,u}^{cl,m}$ & (Un)binding rate of cross-linkers (\textit{cl}) and motor (\textit{m}) & For \textit{cl}, $0.01-10 /s $ \\
  \hline
   $\bar \tau_{b,u}^{cl,m}$ & Mean (un)binding time of cross-linkers (\textit{cl}) and motor (\textit{m}) & -\\
  \hline
$d_s$  & Motor filament step size &$5 \ nm$ \\
\hline
$k_{b,u}^{ms}$& Single motor head (un)binding rate &$0.2$$ / s, 1.7/s$  \cite{Kovacs2007}  \\
\hline
$F_{s}$ & Motor filament stall force&$24 \ pN$  $ ^\textbf{ d}$ \\
\hline
$v_w^{0}$ & Walking velocity of motor filament &$10$  $ nm / s$ $ ^\textbf{ e}$ \\
\hline
$\alpha$ & Stall velocity mechanochemical parameter for motor filament&$0.2$ \cite{Erdmann2013} \\
\hline
$\beta$ & Catch-bond mechanochemical paramter for motor filament &$2$ \cite{Erdmann2013}\\
\hline
$x_{cl}$ & Characteristic slip-length for cross-linker &$0.5$ $ nm$ \cite{Ferrer2008}  \\
\hline
$x_m$  & Characteristic catch-length for single motor head &$1.6$ $ nm$ \cite{Guo2006}  \\
\hline
$K_{ms}$ & Stiffness of single motor head light chain &$0.5$ $pN/nm$ \cite{Walcott2012, Veigel2003a}\\
\hline
$K_{m}$ & Effective spring constant of motor filament  &$ - $ \\
\hline
  $\tau_{lab}$& Timescale of laboratory measurements & $200 s$  \\ 
  \hline
   $\epsilon_{lab}$& Divergence point of steady-state and laboratory measurements & $ - $ \\ 
  \hline
$\rchi$ & Stochastic contraction length of aFD & \textit{Observable} \\
\hline
$\xi $& Stochastic extension length of aFD & \textit{Observable} \\
\hline
$\omega_\rchi $& Contraction frequency of aFD & \textit{Observable} \\
\hline
$\omega_\xi$ & Extension frequency of aFD & \textit{Observable} \\
\hline
$v_{eff}$ & Effective inter-filament velocity of aFD & \textit{Observable} \\
\hline

\end{tabular}
}
\caption{Glossary of variables and model parameters chosen to mimic a typical system of actin filaments, cross-linkers, and non-muscle myosin IIA motor filaments.
\textbf{a} - Geometric constraints of the bipolar motor filament may disallow all tens of heads to be available for binding to a pair of actin filaments due to its double-ended conic structure \cite{Norstrom2010, Billington2013}. Since the number of non-muscle myosin II heads per side of a mini-filament is 30 for isoform A \cite{Billington2013}, we assume for this study that a third of these heads are available for binding to the 1D actin filament.
\textbf{b} - Calculated as the inverse of concentration $V/N$, assuming a bulk cross-linker concentration of 1 $\mu M$.
\textbf{c} - Although the exact volume of a cross-linker binding pocket is unknown, this is an order of magnitude estimate based on the dimensions of $\alpha$-actinin \cite{Wachsstock1993}, which is valid in the approximate expression of Eq.~\ref{eq:kinetics}. Changes of this value would weakly shift the effective stochastic rate constants of cross-linker (un)binding for a given $\epsilon$.
\textbf{d} - An approximation using the stall force of a single myosin II head, $F_{ss} \approx 2 pN$ \cite{Norstrom2010}, multiplied by its duty ratio at stall and the number of motor heads available for binding, $F_{s} = \rho_{s}^m(F) N_t F_{ss} = 12$ $pN$. This is increased to account for two motor head ensembles on each side of the bipolar mini-filament.
\textbf{e} - Calculated using the analytic result of Erdmann et al. \cite{Erdmann2013} for the zero-force walking rate based on individual motor head (un)binding rates. This value was also approximately obtained by Stam et al. \cite{Stam2015} and is further motivated by the experimental results of Norstrom et al. \cite{Norstrom2010}.
}
\label{table:params}
\end{table*}

We now consider the mechanochemical dynamics of the myosin II motor filament included in our model. These highly transient units, as described in the main text, must be regarded as a coarse-grained version of a more detailed stochastic (un)binding and walking process of many individually transient motor heads in a single bipolar mini-filament. We consider an effective unbinding rate of the coarse-grained overlap potential which mimics the mean unbinding time of a single side of the mini-filament with $N_t=10$ heads. Since the overall motor filament unbinding rate $k_{u}^{m}$ is non-trivial compared to the single cross-linker case because of its non-linear dependence on the single motor head (un)binding rates $k_{b}^{ms}$ and $k_{u}^{ms}$ and individual motor head stochasticity, we use an approximate expression for the mean unbinding time of the ensemble of heads derived by Erdmann et al. \cite{Erdmann2013} using a series expansion of an adjoint master equation for motor ensemble states:

\begin{equation}
\label{eq:off0}
\bar \tau_u^{m} \approx \frac{1}{k_{b}^{ms} N_t}\Bigg[\Bigg(1 + \frac{k_b^{ms} }{k_u^{ms}}\Bigg)^{N_t} - 1\Bigg].
\end{equation}

\noindent We note that this form is only accurate under a zero-load assumption, but is remedied by combining with an exponential factor as shown below. In an approximation, we inherently assume a single exponentially-distributed process with rate $k_{u}^{m} = 1 / \bar \tau_u^{cl} $. For simplicity, we also consider motor filament binding to the pair of actin filaments as a single stochastic process with rate $k_{b}^{m} = k_b^{ms} N_t $, ignoring partially bound states which do not generate tension between the actin filament pair.

Individual myosin II motor heads have been shown to display ``catch"-bond behavior, characterized by an increased bound lifetime with applied load \cite{Guo2006, Kovacs2003, Kovacs2007}. We aim to describe the (un)binding and walking kinetics of an ensemble of motor heads with a simple set of parameters to capture the essential aspects of motor filament mechanosensitivity - a binding lifetime that increases exponentially with applied load $F$ and the number of motor heads in the filament $N_t$, motivated by the results of \cite{Erdmann2013}, and a walking velocity that is one of the celebrated Hill form \cite{Hill1939}. Since the exponential dependence of motor head unbinding in the post power-stroke state has been directly observed in the bound lifetime of an entire ensemble of heads \cite{Erdmann2013}, we express the unbinding rate of the motor filament as:

\begin{equation*}
\label{eq:catch}
k_{u,eff}^{m} = k_{u}^{m} \text{exp} \Bigg( \frac{-F x_{m}}{ N_b(F) k_b T} \Bigg),
\end{equation*}

\noindent where $x_{cl}$ is a characteristic unbinding distance of a single motor head, and $N_b(F)$ is the number of motor heads per side of the motor filament bound to actin. An essential characteristic of the myosin II motor filament to capture in this function is a near-linear increase in the number of bound heads with external load \cite{Duke1999, Piazzesi2007, Erdmann2013}. Since this expression is difficult to derive due to the stochastic nature of the motor heads, an approximate expression for $N_b(F)$, motivated by \cite{Erdmann2013}, can be written as $N_b(F) = \rho N_t + \beta F$ where the mechanosensitivity parameter $\beta$ has been chosen to mimic the response of a low-duty ratio motor like non-muscle myosin IIA. Upon a motor stepping event, the number of bound heads determines the effective spring constant of the bipolar mini-filament with $N_t$ heads bound per side to an actin filament in parallel, with stiffness $K_{ms}$ corresponding to the myosin II light chain connecting individual heads. The sides of the bipolar filament are mechanically connected by an extremely stiff motor filament heavy chain region, giving an effective spring constant of the ensemble $K_m = K_{ms} N_t / 2$. 

In the case of motor walking, we can include a force-dependent walking velocity $v_w(F_m)$ similar to a Hill-relation \cite{Hill1939} for myosin II:

\begin{equation*}
\label{eq:hillstall}
v_w(F) = 2 v_w^{0} \frac{F_{s} - F_m}{F_{s} + F / \alpha} 
\end{equation*}

\noindent where $F_m$ is the instantaneous force on the motor, $\alpha$ is a parameter describing the concave nature of the velocity function which has been chosen to mimic a low-duty ratio motor, and the stall force of the motor filament $F_{s}$ is described in Table~\ref{table:params}. This equation accounts for two ensembles of heads walking with velocity $v_w^0$ in opposite directions on each actin filament. We note that in general, non-muscle myosin II filaments are highly transient compared to their muscle sarcomere counterpart, but this still has shown to be an effective representation of stall dynamics \cite{Erdmann2013}. This transience is due to a small number of motor molecules $N_t \approx 28$ \cite{Billington2013} as compared to smooth or skeletal muscle filaments where $N_t \approx 500$ \cite{Piazzesi2007}, causing frequent detachments of the motor head ensemble from the filament pair and tension release, which, in turn, raises serious slippage issues as discussed. 

We have chosen in our stochastic representation to model non-muscle myosin II isoform A, which displays a faster head (un)binding dynamics and less mechanosensitivity compared to isoform B, another abundant motor filament in the eukaryotic cytosol \cite{Kovacs2003}. In \textit{in vitro} sliding assays, myosin II head sliding velocities vary significantly, ranging from 50 $nm/s$ to as slow as 10 $nm/s$ under varying ATP concentration for isoform B \cite{Norstrom2010, Billington2013}. Although to our knowledge, no single-molecule kinetic study of isoform A mini-filaments \textit{in vivo} has been performed - we use a prediction from Erdmann et al. \cite{Erdmann2013} to obtain an average isoform A sliding velocity of $v_w^0 = 10 nm/s$ based on the isoform and number of binding heads we are considering in the mini-filament:

\begin{equation*}
\label{eq:motorwalkingrate}
k_{w}^{m} = ({\rho_{ms}}^{-1} - 1) k_{b}^{ms}
\end{equation*}

\noindent where $\rho_{ms} \approx 0.1$ is the zero-force duty ratio of a single motor head determined experimentally \cite{Kovacs2003}.

\section*{\label{sec:levelAA}Appendix B - Power-law behavior of the passive force dipole}

We here perform a short-time and asymptotic analysis of the contraction equation presented in this paper (Eq~\ref{eq:effcrosslinkermotionavg}). We rewrite the original contraction ODE as:

\begin{equation*}
\label{eq:crosslinksimplify}
\frac{dl_o}{dt} = ae^{-\gamma l_o} - bl_oe^{-\gamma l_o}, 
\end{equation*}

\noindent where $a \equiv \epsilon P_o^{cl} / 2 K_t \eta$, $b \equiv K_t/ \eta$, and $\gamma \equiv ln(c) / \Delta$ where $c \equiv 1+v\text{exp}(\epsilon/k_bT)$. We can also write the predicted steady-state overlap as $l_o^{ss} \equiv \frac{b}{a}$. This equation is now separable and yields:

\begin{equation*}
\label{eq:crosslinksimplify2}
\text{Ei}(-\gamma l_o^{ss} ) - \text{Ei}\Big(\gamma \Big(l_o - l_o^{ss} \Big)\Big) = bte^{-\gamma l_o^{ss}}
\end{equation*}

\noindent where $\text{Ei}(\cdot)$ represents the exponential integral $\text{Ei}(x) = \int_{-x}^{\infty} e^{-x} dx / x$. We can expand in the short time limit, i.e. at $l_o \rightarrow 0$, $ \text{Ei}\Big(\gamma \Big(l_o - l_o^{ss} \Big)\Big) \approx \text{Ei}(-\gamma l_o^{ss} ) - (l_o / l_o^{ss})e^{-\gamma l_o^{ss}} + O(l_o^2)$ to give a linear force-generating regime:

\begin{equation*}
\label{eq:crosslinkshorttime}
l_o \approx \frac{\epsilon P_o^{cl}}{2 \Delta \eta} t.
\end{equation*}

\noindent Thus the predicted sliding velocity of the filaments is $v_o^{cl} = \frac{\epsilon P_o^{cl}}{2 \Delta \eta}$ in absence of kinetic arrest. In the long time limit we can expand around $l_o = l_o^{ss}$ such that approximately $\text{Ei}\Big(\gamma \Big(l_o - l_o^{ss} \Big)\Big) \approx \text{ln}\Big(\gamma \Big(l_o^{ss} - l_o \Big)\Big)$. This immediately gives an exponentially decaying overlap function to steady state:

\begin{equation*}
\label{eq:crosslinklongtime}
l_o \approx l_o^{ss} - \Bigg[ \frac{1}{\gamma} \text{Ei}(-\gamma l_o^{ss}) \Bigg] e^{- b c^{-l_o^{ss}/\Delta} t}
\end{equation*}
\\
\noindent where the timescale of approaching steady state is given as $1/\tau_{ss} = bc^{-l_o^{ss} / \Delta} = \frac{K_t}{\eta} (1 + v \text{exp}(\epsilon / k_bT))^{-l_o^{ss} / \Delta}$. Surprisingly, this can also be related to the simulation trajectories shown in the text with using $l_o(\tau_{lab})$ in substitution for $l_o^{ss}$ when $\epsilon > \epsilon_{lab}$. So, we use $l_o(\tau_{lab})$ in fitting for the glassy regime. To differentiate between the two power law regimes, a transition time $\tau_{trans}$ is chosen when the slope of the time-series in $l_o$ deviates by 10$\%$ for each cross-linker energy $\epsilon$. Based on this analysis, the presence of myosin II motors induces two major changes in the behavior of the element: (1) a short-time regime independent of $\epsilon$ replaces $v_o^{cl} \propto \epsilon$, and (2) a significant decrease in relaxation timescale $\tau_{ss}$ as compared to the passive cross-linker induced glassy state when in absence of motor dynamics. 

\section*{\label{sec:levelAB}Appendix C - Kinetic behavior of the active force dipole}

We can derive the governing equation of the dynamics of the aFD from (un)binding kinetics of the passive and active components. We use the simple kinematics of a contraction event: $\xi_0(l_o,\tau) = l_o(1 - \text{exp}(\frac{-K_t}{2\eta}\tau))$ and extension event $\rchi_0(l_o,l_w,\tau) = ( \frac{K_m}{K_{eff}}(l_o + l_w) - l_o)(1 - \text{exp}(\frac{-K_{eff}}{2\eta}\tau))$ that come directly from solving the overlap relaxation of the element when a motor is (un)bound ($m=0,1$) and cross-linkers are unbound ($n=0$.) We have defined $K_{eff} = K_m + K_t$. If we are considering a cross-linker binding reaction with rate constant $k_r$ in our system which stops the contraction or extension motion, we must average $\rchi$ and $\xi$ over the possible holding times of that reaction, $P(\tau | k_r) = k_r\text{exp}(-k_r \tau)$. Evaluating $\bar \xi(l_o) = \int_{0}^{\infty} \xi_0(l_o,\tau) P(\tau | k_r) d\tau$ and $\bar \rchi(l_o, l_w) = \int_{0}^{\infty} \rchi_0(l_o,l_w,\tau) P(\tau | k_r) d\tau$, we have

\begin{eqnarray*}
\label{eq:kinematicavg}
&  \bar \rchi(l_o,k_r)  = l_o\bigg(1+\frac{2\eta k_r}{K_t}\bigg)^{-1}  , \  \text{and}\\ 
& \bar \xi(l_o,l_w,k_r) =  \bigg( \frac{K_m}{K_{eff}}(l_o + l_w) - l_o\bigg)\bigg(1+\frac{2\eta k_r}{K_{eff}}\bigg)^{-1}.
\end{eqnarray*}

\noindent We must now consider the walk length of the motor filament in a time $\tau$, $l_w = \int_{0}^{\tau} v_w dt$, since it fundamentally depends on the kinetics of the reactions leading up to a contraction event. We consider the motor filament  walking in the $n\neq0$ state such that $F_m \approx K_m l_w$. The motor filament walk length time is controlled by cross-linker unbinding, since we must consider another kinematic process once $n=0$.  We approximate the motor filament stall-force relation in Eq.~\ref{eq:hillstall} as $v_w(F) \approx v_w^{0}(1-K_m l_w / F_s)$. This is simply a linear approximation of the true concave force-velocity relation which accounts for the $F=0$ and $F=F_s$ behavior. Integrating this equation for $l_w$, and averaging over all possible cross-linker unbinding times with rate $k_u^{cl}$ as $\bar l_w(l_o) = \int_{0}^{\infty} l_w(l_o,\tau) P(\tau | k_u^{cl}) d\tau$, we have:

\begin{equation*}
\label{eq:walklength}
\bar l_{w}(l_o) = \frac{F_s}{K_m} \bigg(1+\frac{F_s k_u^{cl}(l_o)}{K_m v_w^0}\bigg)^{-1}.
\end{equation*}

\noindent To write full expressions for $\rchi(l_o)$ and $\xi(l_o)$ as observed in simulation, we must consider the possible series of reaction events and their probabilities when a contraction or extension event takes place. We use a first-moment approximation of the actual distribution of contraction distances by considering the succession of kinetic events possible, their mean behavior and their probabilities of occurrence - because all random variables considered are exponential holding times, we can easily compute probabilities of events happening in succession via the memoryless property of exponential distributions. With this in mind, the expected value of contraction, given an initial overlap $l_o$ can be written as:

\begin{eqnarray*}
\label{eq:ratchetfunction}
\mathbb{E} [\rchi | l_o] \approx && \underbrace{\bigg[\bar \rchi(l_o, k_b^{cl}) + \bar l_{w}(l_o)\bigg]}_{n\rightarrow1} P(\tau_b^{cl} < \tau_u^m) + \\ 
&& \underbrace{\bigg[\bar \xi\bigg(\bar \rchi_{\infty}, k_b^{cl}\bigg)\bigg]}_{m\rightarrow0, \ n\rightarrow1} P(\tau_u^{m} < \tau_b^{cl}) P(\tau_b^{cl} < \tau_b^{m}) + ...
\end{eqnarray*}

\noindent The first term in this equation describes the simple probabilistic pathway where $n \rightarrow 1$ before motor filament unbinding. The latter branches are the various pathways in which the motor filament can become (un)bound before cross-linkers rebind. We have defined $\bar \rchi_{\infty}$ to be the long-time limit of $\bar \rchi(l_o,k_r)$ such that $k_r \rightarrow 0$. Since this pathway is only probable if $P(\tau_u^{m} < \tau_b^{cl})$, we can safely assume that $\bar \rchi$ relaxes fully in this case since $\tau_u^{m} \gg K_m / 2 \eta$. The second branch describes the series of events in which $m \rightarrow 0$ and $n \rightarrow 1$ occur in succession. Other probabilistic pathways are insignificant to the mean behavior of the element. The frequency of these contraction events is simply:

\begin{equation*}
\label{eq:ratchetfreq}
\omega_\rchi(l_o) \approx \bigg( \frac{\bar \tau_b^{cl}(l_o) +  \bar \tau_u^{cl}(l_o)}{\rho_m} \bigg)^{-1}
\end{equation*}

\noindent where we have defined the motor duty ratio in the zero-force limit: $\rho_m = k_u^m / (k_u^m + k_b^m)$. In a similar manner, the expected value of extension given an initial overlap can be written in a first-moment approximation as:

\begin{eqnarray*}
\label{eq:slipfunction}
\mathbb{E} [\xi | l_o] \approx  && \underbrace{\bigg[\bar \xi(l_o, k_b^{cl}) \bigg]}_{n \rightarrow 1} P(\tau_b^{cl} < \tau_b^m) + \\
&& \underbrace{\bigg[\bar \xi(l_o, k_b^{m}) + \bar l_{w}(l_o) \bigg]}_{m \rightarrow 1, \ n \rightarrow 1} P(\tau_b^{m} < \tau_b^{cl}) P(\tau_b^{cl} < \tau_u^{m})  + ...
\end{eqnarray*}

\noindent with a frequency:

\begin{equation*}
\label{eq:slipfreq}
\omega_\xi(l_o) \approx \bigg( \frac{\bar \tau_b^{cl}(l_o) +  \bar \tau_u^{cl}(l_o)}{1-\rho_m} \bigg)  ^{-1}
\end{equation*}

\noindent where the first branch describes $n \rightarrow 1$ before motor filament binding. The second branch then describes the case of $m \rightarrow 1$ and $n \rightarrow 1$ in succession.

\bibliography{library}
\end{document}